\documentclass[journal]{IEEEtran}
\usepackage{cite}
\usepackage{amsmath,amssymb,amsfonts}
\usepackage{algorithmic}
\usepackage{graphicx}
\usepackage{textcomp}
\usepackage{xcolor}
\usepackage{multirow}
\usepackage{array}
\usepackage{threeparttable}
\usepackage{subfigure}
\usepackage{stfloats}
\usepackage{caption}
\usepackage{dutchcal}

\makeatletter

\newcommand{\Rmnum}[1]{\expandafter\@slowromancap\romannumeral #1@}
\makeatother

\def\BibTeX{{\rm B\kern-.05em{\sc i\kern-.025em b}\kern-.08em
    T\kern-.1667em\lower.7ex\hbox{E}\kern-.125emX}}
\begin{document}

\title{An Extended Kalman Filter for Distance Estimation and Power Control in Mobile Molecular Communication}
%
%

\author{
\IEEEauthorblockN{Dongliang Jing$^{1,2}$, Yongzhao Li$^1$$^*$, and Andrew W. Eckford$^2$}

\thanks{This work was supported by  the  National Natural Science Foundation of China under Grant 61771365, Grant 61901333, Grant 61901345, Grant 62001358, and by a Discovery grant from the Natural Sciences and Engineering Research Council.}

\IEEEauthorblockA{$^1$State Key Laboratory of Integrated Services Networks,
Xidian University, China, Xi'an, 710071\\
$^2$ Department of Electrical Engineering and Computer Science, York University, Toronto, Ontario, Canada\\
$^*$Corresponding author: yzli@mail.xidian.edu.cn}
}

\markboth{IEEE Transactions on Communications}%
{Submitted paper}

\maketitle

%
%

\begin{abstract}
In this paper, we consider a mobile molecular communication (MC) system consisting of two mobile nanomachines, a transmitter and a receiver, propelled by a positive drift velocity and Brownian motion in a realistic blood-vessel-type flow regime.
Considering the nonlinear movement of the nanomachines, an extended Kalman filter is employed to estimate the distance from the transmitter. Furthermore, based on the predicted distance, to keep the number of received molecules for bit $1$ at a stable level, we employ power control on the number of transmitted molecules based on the distance between the transmitter and the receiver and the residual molecules in the channel from the previous transmission. Finally, the optimal detection threshold is obtained by minimizing the error probability. It is verified that a fixed optimal detection threshold can be effective for the power control scheme in the mobile MC.
The bit error rate (BER) performance of our scheme is verified via simulation results.
\end{abstract}

%

\section{Introduction}

In molecular communication (MC), molecules are employed as information carriers, where information can be encoded by the concentration of molecules, the type of molecules, or the the time of release of the molecules. Aside from its role in nanonetworking and and potential nanomedical advances \cite{farsad2016comprehensive,akyildiz2019moving}, MC has been employed to model the spread of infectious diseases via aerosols during the COVID-19 pandemic \cite{chen2021resource,khalid2020modeling}.

In many proposed applications of MC, transmitters and receivers are mobile, and are assumed to move using Brownian motion \cite{nakano2019methods,varshney2018flow}, a situation known as  mobile MC. In this direction, previous work has considered systems both with and without flow:
in the former case, work has considered channel modelling \cite{srinivas2012molecular}, mobility in the presence of flow \cite{lin2019concentration}, and analysis of multiple-source interference and inter-symbol interference (ISI) \cite{chouhan2019optimal}; in the latter case, work has analyzed mobile transmitters and receivers \cite{huang2019statistical} and the impact on mobility of channel impulse response \cite{cao2019diffusive}. In all of these  examples, mobile MC presents an extremely challenging communication environment due to the time-varying distance, which governs the system's signal strength and ISI. With knowledge of distance, mitigation techniques can be adopted: various methods have been proposed to mitigate ISI, such as modulation/coding methods \cite{gursoy2021concentration,kislal2019isi}, detection methods \cite{li2019csi, thakur2020iterative, qian2021k}, or both of them\cite{chen2020generalized,tang2020molecular}. In \cite{ahmadzadeh2018stochastic}, a diffusive mobile
MC system is analyzed, where the initial distance at the beginning of the block is used as ``outdated'' channel state information. Power control can also be used as a mitigation strategy for both varying signal strength and ISI \cite{tepekule2015isi,jing2020power}.
In \cite{tepekule2015isi}, to mitigate ISI in static MC, the authors adjust the number of transmitted molecules based on the number of residual molecules in the channel.

In this paper we focus on power control, and the distance estimation that is required to adopt a power control strategy.
Conventional approaches to distance estimation include detecting the peak of the channel impulse response \cite{huang2013distance,wang2015distance}, jointly estimating distance alongside other parameters \cite{lin2016parameter,chen2020parameter}, pilot symbols \cite{huang2020initial} or feedback \cite{moore2012measuring}. Alternatively, nanomachine localization methods are proposed in \cite{turan2018transmitter,liu2019localization,kumar2020nanomachine,yetimoglu2021multiple}, making the position of the nanomachine known, whether that knowledge is exact or noisy.
In \cite{turan2018transmitter}, considering ring-shaped observing receivers, the location of the transmitter is analyzed in a vessel-like environment for diffusion-dominated motion.
In \cite{liu2019localization}, considering the two-dimensional diffusion-based molecular communication, four localization schemes based on trilateration method are proposed in various scenarios.
In \cite{kumar2020nanomachine}, considering a three-dimensional diffusion-based molecular communication system and a transparent, spherical receiver, the transmitter is localized when the locations of the receivers are both known and unknown.

In this paper, we use the assumption of localization, i.e., that the terminals have imperfect knowledge of their own position. In static MC, the distance estimation problem reduces to communication, as each terminal may transmit its information to the other. However, this is not the case in mobile MC, where there are two important complications. First, communicating the distance takes a significant amount of time, and the terminals may move in the meanwhile. Second, the realistic flow velocity may depend on the location of the terminals: in particular, in a blood vessel, nanomachines placed at different distances to the center of the vessel will experience different flow velocities \cite{wicke2018modeling}. This situation is critical for future nanomedical applications.

Addressing these problems, we derive and employ an {\em extended Kalman filter} to predict the distance between the terminals, taking into account the location-dependent flow and noisy knowledge of each terminal's position. We then apply the predicted distance to a power control scheme, which improves the performance of mobile MC by adjusting the signal strength and reducing ISI.
Compared with \cite{tepekule2015isi}, in this paper we control the number of transmitted molecules based on the distance between the transmitter and receiver, which varies with time.
The main contributions of this paper are:
\begin{itemize}

\item We derive and use an extended Kalman filter to predict the distance between the transmitter and receiver, and show that this algorithm has excellent performance in our realistic propagation model. We also show that the computational complexity are is within the expected capabilities of MC systems.

\item Based on the Kalman-predicted distance, we control the number of emitted molecules based on the distance between the transmitter and receiver and the residual molecules in the channel to keep the number of received molecules at a stable level for bit $1$. The optimal detection threshold is derived by minimizing the error probability.

\item We show that a distance-based power control scheme can keep the optimal detection threshold nearly constant with respect to the distance, thereby simplifying detection and improving performance.

\end{itemize}

The remainder of this paper is organized as follows. In Section \uppercase\expandafter{\romannumeral 2}, the system model of the mobile MC system in narrow channels, such as blood vessels, is introduced. In Section \uppercase\expandafter{\romannumeral 3}, the distance  between the mobile transmitter and the receiver is derived. In Section \uppercase\expandafter{\romannumeral 4}, the extended Kalman filter is introduced for distance prediction. In Section \uppercase\expandafter{\romannumeral 5}, the optimal detection threshold is analyzed by minimizing the error probability.
In Section \uppercase\expandafter{\romannumeral 6}, a power control scheme is proposed and the performance of bit error rate (BER) is analyzed. In Section \uppercase\expandafter{\romannumeral 7},
simulation results are given which demonstrate the performance of our extended Kalman filter approach and the power control scheme.

\section{System Model}
\label{sec:model}

\subsection{Physical model}

\begin{figure}[!t]
  \centering
  \includegraphics[width=0.45\textwidth]{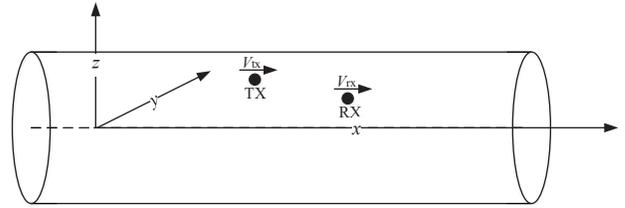}\\
  \caption{The system model of the considered mobile MC system.}\label{Fig2}
\end{figure}

In this paper, we consider a nanomachine-based MC system consisting of a mobile transmitter and mobile receiver inside a 3-dimensional (3D) fluid medium.
As shown in Fig. \ref{Fig2}, we consider a narrow cylindrical tube as our propagation environment; motivated by nanomedical applications, this environment is a simplified representation of a blood vessel. (A true blood vessel would have additional complications, such as obstacles in the form of blood cells, which we do not consider in this paper.)
The medium is assumed to have constant temperature and constant viscosity. The transmitter and receiver, depicted in the figure, are modeled as
spheres with radius $r_{tx}$ and $r_{rx}$, respectively.
To convey the intended information, signal molecules are released from the center of the transmitter and propagate to the receiver via Brownian motion with flow; moreover, at the same time, the transmitter and receiver are mobile, and move within the medium by Brownian motion with flow. The diffusion coefficients of the transmitter, receiver, and signal molecules are $D_{tx}$, $D_{rx}$, and $D_m$, respectively.

We assume the fluid stream in the vessel is free of turbulence, which would be the case for a blood vessel far away from the heart. Therefore, the pressure and velocity at any position can be assumed to be constant over time \cite{felicetti2014molecular,he2016channel}.
%
%
Furthermore, there exists flow only along the axis of the vessel, which we define as the $x$ direction.
Thus, in the
$x$ direction,
the nanomachines are affected by Brownian motion and flow, while in the $y$ direction and $z$ direction, the nanomachines are affected only by Brownian motion.

The velocity profile of the laminar flow was shown to have a parabolic shape, modeled by the well-known Hagen–Poiseuille equation, derived from the Navier-Stokes equations \cite{batchelor2000introduction}:
\begin{align}
\label{eqn:NavierStokes}
{{v \left(r \right)}} = \frac{1}{{4\mu }}\kappa\left( {{R_v^2} - {{ r }^2}} \right),
\end{align}
where $\kappa$ is a constant, which varies with pressure in the per unit length and can be expressed $\kappa = \frac{{\Delta p}}{L}$ (in which $\Delta p$ is the change in pressure along a vessel section of length $L$). Moreover,
$r$ is longitudinal distance of the vessel, $v\left( r\right)$ is the velocity profile, $R_v$ is the radius of the vessel, and $\mu$ is the fluid viscosity.
\begin{figure}[!t]
  \centering
  \includegraphics[width=0.45\textwidth]{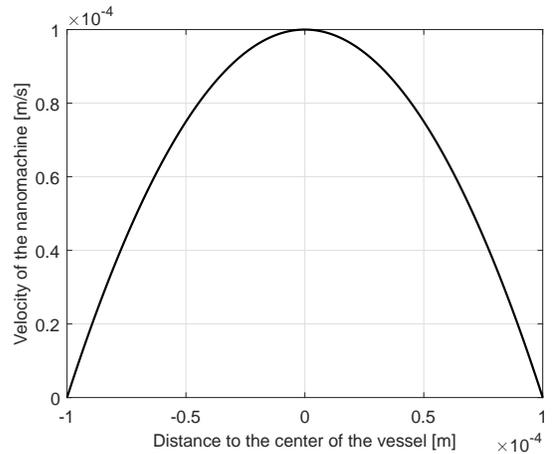}\\
  \caption{The velocity profile varies with the distance to the center of the vessel.}
  \label{Fig3}
\end{figure}
The velocity profile varies with the distance to the center of the vessel, an example of which is shown in Fig. \ref{Fig3}. As can be seen from the figure, the maximum velocity is achieved at the center of the vessel and decreases with increasing distance from the center of the vessel.

\subsection{Communication model}

It is assumed that the mobile transmitter and receiver perform independent random walks with initial locations
%
%
$L_{tx}^0 = ( X_{tx}^0, Y_{tx}^0, Z_{tx}^0 )$, $L_{rx}^0 = ( X_{rx}^0, Y_{rx}^0, Z_{rx}^0 )$ and average drift velocities $v_{tx}^0$, $v_{rx}^0$, respectively. (Note that the drift velocity is a scalar, since the only drift is in the $x$ direction, and that drift velocity is a function of $Y$ and $Z$.) After $k$ steps, the transmitter's location is denoted as $L_{tx}^k = ( X_{tx}^k, Y_{tx}^k, Z_{tx}^k)$ and velocity as $v_{tx}^k$ (and similarly for the receiver, with $rx$ in the subscript).

The transmitter sends information bits $b_{tx} =[b_{tx,1}, b_{tx,2},..., b_{tx,k}]$ to the receiver, where ${b_{tx,k}} \in \{ {0,1} \}$ denotes the information bit in the $k$th bit interval. An on-off keying (OOK) modulation scheme is employed to transmit the information bits: at the beginning of the $k$th bit interval, the transmitter releases either $N_{tx}$ molecules to transmit symbol 1, or zero molecule to transmit symbol 0.

On the receiver side, in this paper we consider a passive receiver which is commonly assumed to be transparent, so that the molecules can enter and leave the receiver via free diffusion. The receiver nanomachine can sense molecules once the molecules arrive in its observation volume, but it does not affect the diffusion of the molecules in any way.  Therefore, the receiver does not affect the flow in the channel. What is more, the effect of the receiver on the flow in the channel is commonly ignored, see e.g. \cite{jamali2019channel}.
To detect a bit, the receiver senses the number of molecules within its observation volume to determine whether bit 0 or bit 1 is transmitted.
If the number exceeds the pre-set threshold, the information is decoded as bit 1; otherwise, bit 0 is decoded.

\section{Distance Analysis between Transmitter and Receiver}

\subsection{Channel model}

In this paper, for simplicity, we consider the dispersion regime described in \cite{wicke2018modeling}, described as follows. Let $v_{{\rm{eff}}}$ be the mean velocity of the fluid, expressed as:
$v_{\rm{eff}} = |\partial_x \mathcal{P}| R_v^2 / (8 \eta)$,
 where ${{\partial _x}\mathcal{P}}$ is the pressure gradient, $R_v$ is the radius of the blood vessel, and $\eta$ is the viscosity of the fluid; furthermore, let $d_x^k$ be the distance between the transmitter and receiver in the $x$ direction during the $k$th step
 \cite{bruus2008theoretical}. Then in the dispersion regime, we make the assumption that $
{{v_{{\rm{eff}}}R_v} \mathord{\left/
 {\vphantom {{v_{{\rm{eff}}}^kR_v} D_m}} \right.
 \kern-\nulldelimiterspace} D_m} \ll {{4d_x^k} \mathord{\left/
 {\vphantom {{4d_x^k} R_v}} \right.
 \kern-\nulldelimiterspace} R_v}$, {\color{black}recalling that $D_m$ is the diffusion coefficient of signal molecules}.

%
%
Under this assumption, released molecules
fully diffuse across the cross section of the vessel
%
%
 while also moving along the $x$ direction by flow and Brownian motion. In this regime,
considering the uniform concentration assumption where the distance between the transmitter and receiver is large relative to the largest dimension of the receiver, the channel impulse response can be expressed as \cite{wicke2018modeling,jamali2019channel}
\begin{align}
\label{eqn:ptk}
P\left( {t,k} \right) = \frac{V_{rx}}{{\pi {R_v^2}}}\frac{1}{{\sqrt {4\pi {D_{{\rm{tot}}}}t} }}\exp \left( { - \frac{{{{\left( {d_x^k - v_{{\rm{eff}}}t} \right)}^2}}}{{4{D_{{\rm{tot}}}}t}}} \right) ,
\end{align}
 where $V_{rx}$ is the volume of the receiver, and $D_{\rm{tot}} = D_{\rm{eff}} + D_{\rm{rx}}$, {\color{black}where $D_{\rm{eff}}$ is the Aris–Taylor effective diffusion coefficient and given by
${D_{\rm{eff}}} = 1 + \frac{1}{{48}}\left( {\frac{{{{\left( {{v_{\rm{eff}}}{R_v}} \right)}^2}}}{D_m}} \right)$}.

The observation probability is the probability of observation of one output molecule at time $t$ at the receiver when the transmitter is stimulated in an impulsive manner at time $t_0 = 0$.
Simplifying the notation, $P_{k-i+1}$ for probability of molecules released at the beginning of $i$th bit interval and observed during the $k$th bit interval ($i<k$) (i.e. $P_1$ denotes the probability of molecules released at the beginning of $i$th bit interval and observed during the same bit interval). As $d_x^k$ is a key parameter in (\ref{eqn:ptk}), in the next section we focus on the analysis of this distance.

 \subsection{Distance between the transmitter and receiver}
In mobile MC, the distance between the transmitter and receiver is an important parameter. The distance varies with time, which also causes the channel impulse response to vary. In this section, we characterize the movement of the nanomachines and focus on the effect on the distance between the transmitter and receiver.

Consider a 3D Brownian motion in discrete time, with flow and Brownian motion in the $x$ direction and only Brownian motion in the $y$ and $z$ directions. At time $k$, the increment in each direction for nanomachine $n$ ($n \in \{tx,rx\}$ for transmitter and receiver, respectively) can be expressed as $\Delta X_{n}^k$, $\Delta Y_{n}^k$, and $\Delta Z_{n}^k$.
In our model, we assume that the motion of the nanomachine in the $y$ and $z$ directions affects the flow velocity in the $x$ direction, so we first analyze the motion of the nanomachines in the $y$ and $z$ directions.
We assume that the Brownian motion is given by a discrete-time Wiener process, with discrete time interval $T$. Considering nanomachine $n$, the increments $\Delta Y_{n}^k$ and $\Delta Z_{n}^k$ are (independent) increments of a discrete-time Wiener process, which follows normal distribution, ${B_{n}^k}\sim{\mathcal N} ( {0, 2D_nT} )$, where $D_n$ is the diffusion coefficient of the nanomachine.
Thus,
\begin{align}
    {Y_n^k} &\sim {\mathcal N}\Big( {{Y_n^0}, 2 k D_n T} \Big),  \\
    {Z_n^k} &\sim {\mathcal N}\Big( {{Z_n^0}, 2 k D_n T} \Big),
\end{align}
%
%
where $Y_n^0$ and $Z_n^0$ are the initial position in the $y$ direction and  $z$ direction, respectively; and where $Y_n^0$ and $Z_n^0$ are independent.
%
%

In the $x$ direction, the nanomachine is affected by Brownian motion as well as flow, where the flow depends on the longitudinal distance of the nanomachine to the center of the vessel. This distance can be expressed as
\begin{align}
\label{eqn:PerpendicularDistance}
d_{n \bot }^k = \sqrt {{{\left( {Y_n^k} \right)}^2} + {{\left( {Z_n^k} \right)}^2}} .
\end{align}
Therefore, the flow velocity in the $x$ direction of nanomachine $n$ during step $k$ can be expressed as
\begin{align}
\label{eqn:vnk}
{{v_n^k}} = \frac{1}{{4\mu }}\kappa\left( {{R_v^2} - {{ (d_{n \bot }^k })^2}} \right)
\end{align}
(see (\ref{eqn:NavierStokes})).
With these definitions,
 the increment $\Delta X_{n}^k$ is composed of a Brownian motion component $B_{n}^k$ and a flow component $V_{n}^k$, with
\begin{align}
\label{equation 1}
    \Delta X_{n}^k = B_{n}^k + V_{n}^k,
\end{align}
where $V_{n}^k = v_n^k T$, and  $v_n^k$ is given by (\ref{eqn:vnk}). The value of $v_n^k$ is assumed to be constant during the $k$ step.
Thus, in the $x$ direction, the position of nanomachine $n$ after $k$ steps, with initial position $X_n^0$, is given by
\begin{align}
\label{equation 2}
\begin{array}{rl}
X_n^k &= X_n^0 + \sum\limits_{i = 1}^k {\Delta X_n^i} \\
&= X_n^{k - 1} + \Delta X_n^k .
\end{array}
\end{align}
After time $k$, the distance between the transmitter and the receiver in the $x$ direction can be expressed as
\begin{align}
d_x^k = \left| {X_{rx}^k - X_{tx}^k} \right|,
\end{align}
where $X_{rx}^k$ and $X_{tx}^k$ are the positions of the receiver and transmitter at the $k$th time slot, respectively, and $X_{rx}^k$ is dependent on the sum of squares in $y$ and $z$ directions.

Our estimation method uses the extended Kalman filter, which is described in detail in the next section. This algorithm requires feedback of the receiver's position, for which we consider two cases:
\begin{itemize}
    \item {\em With feedback:} The position of the receiver nanomachine is encoded and constantly fed back to the transmitter nanomachine using an appropriate feedback method. Such as in \cite{wang2017highly}, a nanomotor with speed over 310 mm/s is considered, which is much larger compared to the drift velocity within the blood vessel, therefore, the latency can be considered to be low. What is more, in \cite{wu2019microrobotic}, a photoacoustic computed tomography method is presented to locate and navigate a nanomotor {\em in vivo} in real time. Thus, an external method can be employed to feed back the positions of the receiver to the transmitter with real time, and prediction can be employed between position measurements. To simplify our analysis, we assume that the transmitter receives feedback information consisting of the receiver's (time-delayed) position.
    \item {\em  With update:} The transmitter is aware of the receiver's initial position, but otherwise uses its own estimates as feedback to the algorithm.
\end{itemize}
It should be clear that the case with feedback is a best-case scenario for the transmitter's knowledge of the receiver's position, while the case with update is a worst-case scenario.
Limited feedback may be available in a practical system: several recent papers have explored feedback of receiver positions \cite{wu2019microrobotic,zhang2019micro,wang2018recent,kim2018artificial,yan2017multifunctional},  or one-way feedback using kinesins walking on microtubules \cite{moore2008molecular,enomoto2011design}, inspired by transport of cargo in eukaryotic cells \cite{jiang2019microtubule} (for a related experimental technique, see \cite{jeune2010cargo}). With limited, irregular feedback, the performance of an actual system will be between the two feedback assumptions given above.

\section{Extended Kalman filter for distance prediction}

As noted in the introduction, the terminals have a noisy estimate of their own positions in space, and can communicate with each other to estimate their distance. However, as described in Section III, the distance changes with time, and the flow rate depends on the position of each terminal in space. Thus, we need to use the out-of-date observations to {\em predict} the actual positions of the receiver nanomachine. These previous positions represent the state variables in our prediction algorithm.

Considering (2), in this section, we mainly focus on the distance estimation in the $x$ direction $d_x^k$.
Therefore, the position of the receiver nanomachine is predicted and employed to estimate the distance $d_x^k$. As positions in the $x$ direction are based on a quadratic equation in the $y$ and $z$ coordinates (see (\ref{eqn:PerpendicularDistance})-(\ref{equation 1})), the system is nonlinear. Such a problem is ideally suited to an {\em extended Kalman filter}, which can be applied to nonlinear problems by locally making a linear approximation algorithm that has been shown to be effective for the state sequences of dynamic systems \cite{reif1999stochastic,daum2005nonlinear}. The importance of using an extended Kalman filter is that our framework can be generalized to include more complex features of motion, such as controlled nanomachine motion and viscous friction, which can be investigated in future work.

The receiver's state at time $k$ is its location $L_{rx}^k$, that is,
\begin{align}
L_{rx}^k = \left[ {\begin{array}{*{20}{c}}
{X_{rx}^k}\\
{Y_{rx}^k}\\
{Z_{rx}^k}
\end{array}} \right] ,
\end{align}
consistent with our notation in the previous section.
The position of the receiver nanomachine based on the system model in the $x$ direction can be expressed in a difference equation as
\begin{align}
X_{rx}^k &= X_{rx}^{k-1} + v_{rx}^k T + B_{x}^k\\
\label{eqn:motion-model-1}
&= X_{rx}^{k-1} + \frac{1}{{4\mu }}\kappa \Big(R_v^2 - (Y_{rx}^{k-1})^2 - (Z_{rx}^{k-1})^2\Big)T + B_{rx,x}^k,
\end{align}
where (\ref{eqn:motion-model-1}) follows from (\ref{eqn:PerpendicularDistance})-(\ref{eqn:vnk}). In the $y$ and $z$ directions, as there is no drift, the position is much more simply expressed as
\begin{align}
    \label{eqn:motion-model-2}
    Y_{rx}^k &= Y_{rx}^{k-1} + B_{rx,y}^k,\\
    \label{eqn:motion-model-3}
    Z_{rx}^k &= Z_{rx}^{k-1} + B_{rx,z}^k .
\end{align}
From these difference equations, it is important to note that, firstly, the next location $X_{rx}^k$ is in a dynamical relationship with each coordinate in $L_{rx}^{k-1} = [X_{rx}^{k-1},Y_{rx}^{k-1},Z_{rx}^{k-1}]^T$, and secondly, that this relationship is quadratic. These features are appropriate for the use of an extended Kalman filter. It is also clear that the process noise $w\left(k\right)$
is given by the Brownian motion $[B_{rx,x}^k,B_{rx,y}^k,B_{rx,z}^k]$, which is a Gaussian random vector for each $k$, with covariance matrix $Q = \sqrt{2D_{rx}T }I$, where $I$ is the identity matrix. Therefore, the extended Kalman filter estimates the state of the considered  mobile molecular communication system modeled by the discrete-time state equation
\begin{align}
\label{eqn:motion-model-4}
L_{rx}^k = f\left( {{L_{rx}^{k - 1}}} \right) + {w(k)},
\end{align}
where $f$ is the state transition function which evolves the state given the previous state $L_{rx}^{k - 1}$.

{\em With feedback:} The observation models are used to predict the next state of the extended Kalman filter and can be expressed as:
\begin{align}
\label{equ 9}
{{\cal Z}}\left( k \right){\rm{ }} = H L_{rx}(k) + n_n(k) ,
\end{align}
where $n_n(k)$ is observation noise (for example, the position sensed by the receiver is noisy and inaccurate).
Since only the motion is observable, the observation model $H$
is the identity matrix, and the
observation is then expressed simply as ${{\cal Z}}\left( k \right){\rm{ }} = L_{rx}(k) + n_n(k)$.
The observation noise  $n_n( k )$ is an IID Gaussian process with diagonal covariance matrix $R$, $n_n(k)\sim \mathcal{N}(0,R)$.


Deriving the extended Kalman filter in the special case of our problem, from (\ref{eqn:motion-model-1})-(\ref{eqn:motion-model-4}),
the state update is given by a vector equation
\begin{align}
    \label{eqn:f}
    f(L_{rx}^{k-1}) &=
    \left[
        \begin{array}{c}
            f_x(L_{rx}^{k-1})\\
            f_y(L_{rx}^{k-1})\\
            f_z(L_{rx}^{k-1})
        \end{array}
    \right]\\
    &=
    \left[
        \begin{array}{c}
            X_{rx}^{k-1} + \frac{1}{4\mu} \kappa (R_v^2 - (Y_{rx}^{k-1})^2 - (Z_{rx}^{k-1})^2)T\\
            Y_{rx}^{k-1}\\
            Z_{rx}^{k-1}
        \end{array}
    \right] .
\end{align}
The
state transition matrix of the extended Kalman filter is the Jacobian matrix of $f$:
\begin{align}
    \nonumber \lefteqn{F(k-1)} \\
    &= \left[ \begin{array}{ccc}
        1 & \partial f_x / \partial Y_{rx} & \partial f_x / \partial Z_{rx} \\
        0 & 1 & 0 \\
        0 & 0 & 1
    \end{array} \right] \\
    &= \left[ \begin{array}{ccc}
        1 &
        - \frac{1}{2 \mu} \kappa Y_{rx}^{k-1}T &
        - \frac{1}{2 \mu} \kappa Z_{rx}^{k-1}T \\
        0 & 1 & 0 \\
        0 & 0 & 1
    \end{array} \right] .
\end{align}
The Kalman filter is implemented through a pair of iterations, one on covariance matrices (used to calculate the Kalman gain), and one on the estimates and predictions themselves. Each iteration has a {\em predict} and an {\em update} step. These are described below.

For the first iteration, define a pair of iteratively updating covariance matrices: and $\tilde P(k|k-1)$, the predicted covariance matrix, and $\tilde{P}(k|k)$, the state covariance matrix. What's more, $\tilde P(k|k-1)$ associate to the Jacobian matrix of the state
transition function $F(k)$, the state covariance matrix $\tilde P(k-1|k-1)$, and process noise covariance matrix $Q$. Supposing $\tilde P(k-1 | k-1)$ is given, the {\em predict} step is given by
\begin{align}
\label{equation 23}
\tilde{P}\left( {k|k-1} \right) &= F\left( k \right)\tilde P\left( k-1|k-1 \right)F{\left( k \right)^T} + Q .
\end{align}
This covariance matrix is used to calculate the Kalman gain $K(k)$, given by
\begin{align}
\label{equation 22}
{K}\left( {k} \right) &= \tilde{P}\left( {k|k-1} \right){\left[ {\tilde{P}\left( {k|k-1} \right) + R} \right]^{ - 1}}.
\end{align}
Next, the {\em update} step is given by
\begin{align}
\label{equation 24}
\tilde{P}\left( {k|k} \right) &= \left( {I - {K}\left( {k} \right)} \right)\tilde{P}\left( {k|k-1} \right),
\end{align}
which can then be fed back to (\ref{equation 23}) to complete the iteration on these covariance matrices.

For the second iteration, we also define a corresponding pair of quantities related to the underlying position of the nanomachine: the state prediction $\hat{L}_{rx}^{k|k-1}$, and the state estimate $\hat{L}_{rx}^{k|k}$. These iterations use the Kalman gain $K(k)$, obtained from the first iteration. Given
$\hat{L}_{rx}^{k-1|k-1}$, the {\em predict} step is given by
\begin{align}
    \label{eqn:L-predict}
    \hat{L}_{rx}^{k|k-1} &= f\Big(\hat{L}_{rx}^{k-1|k-1}\Big) ,
\end{align}
with $f(\cdot)$ given by (\ref{eqn:f}),
and the {\em update} step is given by
\begin{align}
\label{equ 11}
    \hat{L}_{rx}^{k|k} &= \hat{L}_{rx}^{k|k-1} + K(k) \Big[ {\cal Z}_n(k) - \hat{L}_{rx}^{k|k-1} \Big],
\end{align}
recalling the observation model ${\cal Z}_n$.
This iteration is then fed back to (\ref{eqn:L-predict}).

To complete the specification of the iterative algorithm, we must specify the initial estimate for $\hat{L}_{rx}^{0|0}$ and the initial covariance matrix $\tilde P(0|0)$. These are given by $\hat{L}_{rx}^{0|0} = E[L_{rx}^0]$, the initial expected position, and $\tilde P(0|0) = Q$, the covariance of the Brownian motion.

The distance prediction is from the transmitter side and we assume the position of the nanomachine
is known to itself. Based on the position in the $x$ direction,
we can get the predict distance $d_{px}^k$ between the transmitter and the receiver based on the system model:
\begin{align}
\label{kalman_result}
{d_{px}^k} = \left|{ {{{\mathcal{F}}}\hat L_{rx}^{k|k} - {\mathcal{F}} L_{tx}^{k}} }\right|,
\end{align}
where $\mathcal{F}= \left[ {1,0,0} \right]$.

{\em With update:} When there is no physical feedback signal, the transmitter employs the predicted positions $\hat{L}_{rx}^{k|k-1}$ in (24) as the observed states $L_{rx}(k)$ of the Kalman filter in (16). Therefore, the observation model in (16) is modified to
\begin{align}
\label{equ 9b}
{{\cal Z}}\left( k \right) = H \hat{L}_{rx}^{k|k-1} + n_n(k) .
\end{align}
The other steps are the same as the extended Kalman filter with feedback.

In the analysis of the computational complexity of the extended Kalman filter, we consider the number of multiplication (Mult.) and addition (Add.) operations performed during the estimation process. For the 3D predicted state and 3D observation vector, the computational complexity of the extended Kalman filter in terms of these operations is given in Table I.
\begin{table}[h!]
  \begin{center}
    \caption{Computational complexity.}
    \begin{tabular}{l|c|c} 
      \textbf{Instruction} & \textbf{Mult.}& \textbf{Add.}\\
      \hline
       Predicted covariance (Eq.21)  & 54 & 45\\
      \hline
      Kalman gain (Eq.22) & 63 & 36\\
      \hline
      Updated covariance (Eq.23) & 27 & 27\\
      \hline
      Predicted state (Eq.24) & 3 & 3 \\
      \hline
      Updated state (Eq.25) & 9 & 12\\
      \hline
      Total & 156 & 123\\
      \hline
    \end{tabular}
  \end{center}
\end{table}

In terms of the computational complexity of the extended Kalman filter scheme, our method is comparable to and even lower than other algorithms proposed in the molecular communication literature. For example, filter methods are also proposed in DBMC. Wiener and extended Kalman filter detection methods are proposed for DBMC in \cite{9691339}.
To target tumors in the body, in \cite{shi2020nanorobots}, a computationally complex gravitational search algorithm and particle swarm optimization algorithm are proposed to locate the tumor.
Based on \cite{li2020progress} and the rapid development of nanotechnology, it is reasonable to believe that nanomachines can perform moderately complex functions such as the one we propose.

We use the sample mean of the error magnitude in order to evaluate and compare the performance of distance estimation.
This is given by
\begin{align}
\label{equation 28}
e= \frac{1}{k}\sum\limits_{j = 1}^k {\left| {{e_{j}}} \right|}, 
\end{align}
where $e_{j}$ is the error in the $j$th time slot. For the extended Kalman filter prediction, $e_j={d_{px}^j}-d_x^j$, where $d_x^j$ is the actual distance between the transmitter and receiver in the $x$ direction during the $j$th step.

\section{Power Control and Performance analysis}

Using our distance prediction results, here we derive a power control scheme and the optimal detection threshold, which together ensure high performance for mobile MC systems. We first describe a power control scheme to maintain a constant number of received molecules for transmitted bit $1$ even as distance changes. Then, based on the statistical properties of the received molecules and hypothesis testing method, we derive the optimal threshold, and finally we conduct a probability of error analysis for the system.
\subsection{Power control}
As the number of received molecules depends on distance from transmitter to receiver, the varying distance in mobile MC has a significant impact on performance.
A strategy to counteract this effect is to employ {\em power control}, i.e., to stabilize the {\em average} number of molecules at the receiver $\hat{N}_{rx,k}$, by manipulating the number of transmitted molecules $N_{tx,k}$. Here we derive a power control scheme based on the distance estimates provided by our extended Kalman filter.

We first obtain $\hat{N}_{rx,k}$.
In the $k$th time slot, after $N_{tx,k}$ molecules are released by the transmitter, the total number of molecules measured by the receiver can be expressed as
\begin{align}
\label{equation 27A}
{\hat{N}_{rx,k}} &= \sum\limits_{i = 1}^{k} {{N_{tx,i}}{P_{k - i+1}}} \\
\label{equation 27A2}
&= N_{tx,k} P_1 + \sum\limits_{i = 1}^{k-1} {{N_{tx,i}}{P_{k - i+1}}} .
\end{align}
In (\ref{equation 27A2}), the calculation of $\hat{N}_{rx,k}$ is separated into two terms: the first part is the number of molecules transmitted at the beginning of the current time slot and received during the current time slot, while the second part is the number of molecules received in the current time slot but transmitted from the previous time slots (i.e., ISI). In (\ref{equation 27A})-(\ref{equation 27A2}), note that the arrival probabilities $P_{k-i+1}$ vary with the distance between the transmitter and the receiver.

Power control can be performed at each time instant $k$ by solving (\ref{equation 27A2}) for $N_{tx,k}$. We can write
\begin{align}
    \label{equation 55}
    {N_{tx,k}} = \frac{1}{{{P_1}}}\left( {{\hat{N}_{rx,k}} - \sum\limits_{i = 1}^{k - 1} {{N_{tx,i}}{P_{k - i + 1}}} } \right).
\end{align}
In the power control scheme, based on (\ref{equation 55}), $\hat{N}_{rx,k}$ is now the {\em target} average number of received molecules (set as a system parameter); $P_{k-i+1}$ is the transmitter's estimate of arrival probability $k-i+1$ time instants after transmission, which is a function of the transmitter's distance estimates; and $N_{tx,i}$ is the number of molecules transmitted at time $i$. All the quantities on the right side of (\ref{equation 55}) are available to the transmitter, as are the distance estimates from the extended Kalman filter.

\subsection{Statistical properties of the received molecules}
To analyze the optimal detection threshold and the system performance, in this subsection, we work on the statistical properties of the received molecules $N_{rx,k}$.

In the $k$th time slot, after $N_{tx,k}$ molecules which controlled by the transmitter based on (\ref{equation 55}) are released by the transmitter, the statistical properties of molecules observed by the receiver can be expressed as
\begin{align}
\label{equation 29}
 {N_{rx,k}} = N_{rx,k}^c + N_{rx,k}^I+ N_{rx,k}^n,
\end{align}
where $N_{rx,k}^c$ is the number of molecules transmitted at the beginning of the current time slot and received during the current time slot, $N_{rx,k}^I$ is the number of molecules received in the current time slot but transmitted from the previous time slots (i.e., ISI), and $N_{rx,k}^n$ is the counting noise. Mathematical models for each quantity are given below.

Let $N_{tx,k}$ information molecules released by the transmitter at the beginning of $k$th time slot and received by the receiver during the current time slot follow a binomial distribution \cite{kuran2010energy}, given as
\begin{align}
\label{equation 30}
N_{rx,k}^c \sim \text{Binomial}\left( {{N_{tx,k}},{P_1}} \right) .
\end{align}
Considering that the number of molecules $N_{tx,k}$ is large, the binomial distribution can be approximated by a normal distribution, and can be expressed as
\begin{align}
\label{equation 31}
 N_{rx,k}^c \sim \mathcal{N}\left( {{N_{tx,k}}{P_1},{N_{tx,k}}{P_1}(1 - {P_1})} \right)  ,
\end{align}
The ISI $N_{rx,k}^I$ accounts for the molecules transmitted from previous time slot and received during the current time slot, expressed as
\begin{align}
\label{equation 32}
 N_{rx,k}^I = \sum\limits_{i = 1}^{k - 1} {\mathbb {N}_{rx,i}^I}  ,
\end{align}
where
\begin{equation}
\label{equation 33}
  \mathbb {N}_{rx,i}^I \sim \mathcal{N}\left( {{N_{tx,{i}}}b_ { i}{P_{k-i+1}},{N_{tx,i}}b_{i}{P_{k-i+1}}(1 - {P_{k-i+1}})} \right) .
\end{equation}
%
The particle counting noise $N_{rx,k}^n$ is a
random process representing fluctuations in the measured concentration due to single events of particles entering/leaving the receptor space \cite{pierobon2011diffusion}. To determine an appropriate value, we follow methods given in \cite{kilinc2013receiver}: the counting noise is assumed to follow a Gaussian distribution $N_{rx,k}^n \sim (\mu_{n,k}, \sigma_{n,k}^2)$ with mean $\mu_{n,k} = 0$ and variance $\sigma_{n,k}^2$ which is dependent on the expected number of molecules received by the receiver, and can be expressed as $\sigma_{n,k}^2 = { N_{rx,k}}/{V_{rx}}$, where $N_{rx,k}$ is the number of received molecules during the $k$th bit interval, and $V_{rx}$ is the volume of the receiver.
\subsection{Hypothesis testing problem and optimal threshold}
Symbol detection can be formulated as the binary hypothesis testing problem
%
%
\begin{align}
\label{equation 34}
  \begin{array}{l}
{H_0}:{N_{rx,k}} = N_{rx,k}^I + N_{rx,k}^n,\\
{H_1}:{N_{rx,k}} = N_{rx,k}^c + N_{rx,k}^I + N_{rx,k}^n,
\end{array}
\end{align}
where $H_0$ and $H_1$ denote the null and alternative hypothesis corresponding to the transmission of bit 0 and 1, respectively, during the $k$th time slot. Considering that the received molecules $N_{rx,k}^c$, $N_{rx,k}^I$ and $N_{rx,k}^n$ are assumed to be independent and follow the normal distribution, their sum $N_{rx,k}$ also follows the normal distribution.
%
%
Therefore, $N_{rx,k}$ for the different hypotheses are distributed as
\begin{align}
\label{equation 35}
 \begin{array}{l}
{H_0}:{N_{rx,k}}\sim{\mathcal{N}}({\mu _{0,k}},\sigma _{0,k}^2),\\
{H_1}:{N_{rx,k}}\sim{\mathcal{N}}({\mu _{1,k}},\sigma _{1,k}^2),
\end{array}
\end{align}
where $\mu_{0,k}$ and $\sigma _{0,k}^2$ represent the mean and variance of the received molecules, respectively, under hypothesis $H_0$, and $\mu_{1,k}$ and $\sigma _{1,k}^2$ represent the mean and variance of the received molecules, respectively, under hypothesis $H_1$. In this paper, we assume the same probability to transmit bit 0 and bit 1, therefore, the mean $\mu_{0,k}$, $\mu_{1,k}$ and variance $\sigma_{0,k}^2$, $\sigma_{1,k}^2$ at time $k$ under hypothesis $H_0$ and $H_1$ can be expressed as:
%
%
\begin{align}
\label{equation 36}
 {\mu _{0,k}} &= {\mu _{I,k}} + {\mu _{n,k}} = \frac{1}{2}\sum\limits_{i = 1}^{k - 1} {{N_{tx,i}}{P_{k-i+1}}},   \\
  \sigma _{0,k}^2 &= \sum\limits_{i = 1}^{k - 1} {\sigma _{I,j}^2}  + \sigma _{n,k}^2\\ 
&= \sum\limits_{i = 1}^{k - 1} {\left[ {\frac{1}{2} {{N_{tx, i}}{P_{k-i+1}}\left( {1 - {P_{k-i+1}}} \right)} + \frac{1}{4}{{\left( {{N_{tx,i}}{P_{k-i+1}}} \right)}^2}} \right]}\\
&+ {\mu _{0,k}}, \notag\\
%
  {\mu _{1,k}} &= {\mu _{c,k}} + {\mu _{I,k}} + {\mu _{n,k}}\\ 
  &= {N_{tx,k}}{P_1} + \frac{1}{2}\sum\limits_{i = 1}^{k - 1} {{N_{tx, i}}{P_{k-i+1}}},  \\
  \sigma _{1,k}^2 &= \sigma _{c,k}^2 + \sigma _{I,k}^2 + \sigma _{n,k}^2 \\
  &= {N_{tx,k}}{P_1}\left( {1 - {P_1}} \right) + \sum\limits_{i = 1}^{k - 1} {\left[ {\frac{1}{2} {{N_{tx,i}}{P_{k-i+1}}\left( {1 - {P_{k-i+1}}} \right)}} \right]}\nonumber \\
 &\:\:\:\:+ \sum\limits_{i = 1}^{k - 1} {\left[ {\frac{1}{4}{{\left( {{N_{tx,i}}{P_{k-i+1}}} \right)}^2}} \right]}  + {\mu _{1,k}}.
\end{align}
%
%
The full derivation of the mean and variance of the $\mu_{0,k}$, $\mu_{1,k}$, $\sigma_{0,k}^2$ and $\sigma_{1,k}^2$ are shown in Appendix A.

A threshold may be used to distinguish $H_0$ and $H_1$; here we derive the optimal threshold. At the receiver, the symbol detected in the $k$th time slot is:
\begin{align}
b_{rx,k} &= \left\{
\begin{array}{cl}
    0, & N_{rx,k} \geq N_{th} ,\\
    1, & \text{otherwise,}
\end{array}
\right.
\end{align}
where ${{N}_{th}}$ is a preset detection threshold. As the ${b_{rx,k}}$ corresponds to either ${b_{tx,k}}=0$ or ${b_{tx,k}}=1$. Therefore, the hypothesis $H_0$ and $H_1$ for ${N_{rx,k}}$ can be expressed as:
\begin{align}
{{H_0}:}&{\mathcal{f}\left( {{N_{rx,k}}|{N_{tx,k}} = 0} \right)},\\
{{H_1}:}&{\mathcal{f}\left( {{N_{rx,k}}|{N_{tx,k}} = 1} \right)}.
\end{align}
So the false alarm probability $\mathcal{P}_F$ and detection probability $\mathcal{P}_D$ can be expressed as
\begin{align}
&{\mathcal{P}_F} = {\text{Pr}}\left( {{N_{rx,k}} > {N_{th}}|{b_{tx,k}} = 0} \right){\rm{ = Q}}\left( {\sqrt {\frac{{{{\left( {{N_{th}} - {\mu _{0,k}}} \right)}^2}}}{{\sigma _{0,k}^2}}} } \right),\\
&{\mathcal{P}_D} = {\text{Pr}}\left( {{N_{rx,k}} > {N_{th}}|{b_{tx,k}} = {\rm{1}}} \right) = {\rm{Q}}\left( {\sqrt {\frac{{{{\left( {{N_{th}} - {\mu _{1,k}}} \right)}^2}}}{{\sigma _{1,k}^2}}} } \right),
\end{align}
where ${\mu _{0,k}}$, ${\mu _{1,k}} $, $\sigma _{0,k}^2$ and
$\sigma _{1,k}^2 $ are given from (38) - (44). Based on the false alarm probability $\mathcal{P}_F$ and detection probability $\mathcal{P}_D$, the detection error probability $P_e$ can be expressed as
\begin{align}
\label{equ43}
{P_e} =  { {\rm{Pr}}{\left( {b_{tx,k}}=0\right)}}{\mathcal{P}_F} + {\rm{Pr}}{\left( {b_{tx,k}}=1\right)} \left( {1 - {\mathcal{P}_D}} \right) ,
\end{align}
where ${\rm{Pr}}{\left( {b_{tx,k}}\right)}$ is the probability to transmit bit $0$ or $1$. By minimizing $P_e$, the optimal detection threshold can be expressed as
\begin{align}
{N_{th,opt}} = \mathop {\min }\limits_{{N_{th}}} \left\{ {{P_e}} \right\}.
\end{align}
Assuming bits 0 and 1 are equiprobable (i.e., ${\rm{Pr}}{\left( {b_{tx,k}}=0\right)} = {\rm{Pr}}{\left( {b_{tx,k}}=1\right)} = \frac{1}{2}$),
the optimal detection threshold $N_{th,opt}$ can be expressed as
\begin{align}
\label{equ45a}
{N_{th,opt}} &= \frac{{{\mu _{0,k}}\sigma _{1,k}^2 - {\mu _{1,k}}\sigma _{0,k}^2}}{{\sigma _{1,k}^2 - \sigma _{0,k}^2}} \notag\\
&+ \frac{{\sqrt {\left( {{\mu _{0,k}}\sigma _{1,k}^2{\rm{ - }}{\mu _{1,k}}\sigma _{0,k}^2} \right){\rm{ - }}\left( {\sigma _{1,k}^2 - \sigma _{0,k}^2} \right){\cal C}} }}{{\sigma _{1,k}^2 - \sigma _{0,k}^2}},
\end{align}
where
${\mathcal{C}} = \left( {{\mu _{0,k}}\sigma _{1,k}^2{\rm{ - }}{\mu _{1,k}}\sigma _{0,k}^2} \right) - 2\sigma _{0,k}^2\sigma _{1,k}^2 \ln \frac{{{\sigma _{0,k}}}}{{{\sigma _{1,k}}}}$.
The existence of optimal detection threshold and its derivation are shown in Appendix B.

\subsection{Probability of error analysis}
In this section, we analyze the average bit error probability considering our power control and symbol detection scheme.

An error occurs if
${b_{rx,k}} \ne {b_{tx,k}}$, where ${b_{rx,k}}$ denotes the bit received in the $k$th time slot, and ${b_{tx,k}}$ denotes the bit transmitted at the beginning of the same time slot.
As in the power control scheme, the number of transmitted molecules in the $k$th time slot based on the predicted distance and molecules transmitted from the previous time slots. Therefore, the average number of received molecules during the $k$th time slot dependent on the current and previous transmitted bits,
and in full generality, the probability of error may be written
\begin{align}
\begin{split}
\label{ber_average}
{P_e} &= \sum\limits_{{(b_{tx,1}}, \ldots ,{b_{tx,k}}) \in \left\{ {0,1} \right\}^k} \mathrm{Pr} \left( b_{tx,1}, \ldots ,b_{tx,k} \right)\\
& \cdot \mathrm{Pr}\left( \mathrm{error} \:|\:  b_{tx,1}, \ldots ,b_{tx,k} \right) .
\end{split}
\end{align}
Note that the number of terms in this sum is exponential in $k$.
In practice, due to the long duration of the impulse response in MC channels, the probability of error for the current bit can meaningfully depend on dozens or even hundreds of previous bits. Thus, exact calculation of (\ref{ber_average}) may be computationally infeasible.

To estimate the probability of error, we instead use an approximation that takes advantage of our power control scheme, which stabilizes the the average number of received molecules. To start with, we expand (\ref{ber_average}) as
\begin{align}
\begin{split}
\label{ber_average2}
{P_e} &= \sum\limits_{{b_{tx,1}}, \ldots ,{b_{tx,k-1}} \in \left\{ {0,1} \right\}} {{\rm{Pr}}}\left( {{b_{tx,1}}, \ldots ,{b_{tx,k-1},b_{tx,k} = 1}} \right)\\
& \cdot {{\rm{Pr}}}\left( {{{N}_{rx,k}}\left( {{b_{tx,1}}, \ldots ,{b_{tx,k}=1}} \right) \le {N_{th,opt}}} \right)\\
&+ \sum\limits_{{b_{tx,1}}, \ldots ,{b_{tx,k - 1}} \in \left\{ {0,1} \right\}} {{\rm{Pr}}}\left( {{b_{tx,1}}, \ldots ,{b_{tx,k-1}},b_{tx,k} = 0} \right)\\ 
& \cdot {{\rm{Pr}}}\left( {{{N}_{rx,k}}\left( {{b_{tx,1}}, \ldots ,{b_{tx,k}=0}} \right) > {N_{th,opt}}} \right),
\end{split}
\end{align}
where ${N}_{rx,k}\left( {b_{tx,1}}, \ldots ,{b_{tx,k}}\right)$ is the random variable reperesenting the number of received molecules, given transmitted bits $1$ through $k$; the terms differ in that the final decision is performed on $b_{tx,k}$, comparing with threshold $N_{th,opt}$.

Now, since power control will keep the mean number of received molecules constant, we make two approximations: first, we replace ${N}_{rx,k}\left( {b_{tx,1}}, \ldots ,{b_{tx,k}}\right)$ by  its average $\hat{N}_{rx,k}$, effectively taking the average inside the sum. Thus, we have
\begin{align}
\begin{split}
{P_e} & \approx \frac{1}{k}\sum\limits_{i = 1}^k \left[ {{\rm{Pr}}}\left( {{b_{tx,i}}} =1\right){{\rm{Pr}}}\left( {{{\hat N}_{rx,i}} \le {N_{th,opt}}|{b_{tx,i}} = 1} \right)  \right.\\
&\left.+{{\rm{Pr}}}\left( {{b_{tx,i}}}=0 \right){{\rm{Pr}}}\left( {{{\hat N}_{rx,i}} > {N_{th,opt}}|{b_{tx,i}} = 0} \right) \right] \\
&= \frac{1}{k}\sum\limits_{i = 1}^k {\left[ {{{\rm{Pr}}}\left( {{b_{tx,i}}}=1 \right)\left( {1 - {\mathcal{P}_{D,i}}} \right) + {{\rm{Pr}}}\left( {{b_{tx,i}}}=0 \right){\mathcal{P}_{F,i}}} \right]}.
\end{split}
\end{align}
where
${\mathcal{P}_{D,i}}$ and ${\mathcal{P}_{F,i}}$ denote the probability of detection and false alarm at the receiver in the $i$th time slot, respectively. Secondly, we approximate ${\mathcal{P}_{D,i}}$ and ${\mathcal{P}_{F,i}}$ as arising from a Gaussian random variable (a reasonable and widely-used approximation since the number of molecules is large,  even if the underlying random variables are not IID), so that these quantities can be written as
\begin{align}
{\mathcal{P}_{D,i}} = Q\left( {\frac{{{N_{th,opt}} - {\mu _{1,i}}}}{{{\sigma _{1,i}}}}} \right),
\end{align}
\begin{align}
{\mathcal{P}_{F,i}} = Q\left( {\frac{{{N_{th,opt}} - {\mu _{0,i}}}}{{{\sigma _{0,i}}}}} \right).
\end{align}
Thus, the average bit error probability over the $k$ time slots for the same probability to transmit bit $0$ and bit $1$ can be expressed as
\begin{align}
\begin{split}
{P_e} &\approx \frac{1}{k}\sum\limits_{i = 1}^k \left[\frac{1}{2}\left( {1 - Q\left( {\frac{{{N_{th,opt}} - {\mu _{1,i}}}}{{{\sigma _{1,i}}}}} \right)} \right) \right.\\
&\left. + \frac{1}{2}Q\left( {\frac{{{N_{th,opt}} - {\mu _{0,i}}}}{{{\sigma _{0,i}}}}} \right) \right].
\end{split}
\end{align}
In our simulation results, we show that this approximate calculation of $P_e$ is highly accurate, and nearly indistinguishable from simulation results.

\section{Numerical and Simulation Results}
In this section, numerical and simulation results are presented to verify our distance prediction and power control schemes, using the channel model discussed in Section \ref{sec:model}.
The diffusion process is divided into discrete time slots of duration $T$, and
%
simulation parameters are listed in Table \ref{tab:parameters},
except the initial positions of the nanomachines, which are given in the captions of each figure.

\begin{table}
\begin{center}
\caption{Simulation parameters}
\label{tab:parameters}

\begin{tabular}{ |c|c|c| }
\hline
Parameters & Symbol & Values  \\
\hline
Fluid viscosity & $\mu$ & 1.3 mPa $\times$ s\\
\hline
Vessel radius & $R_v$ & 10 $\mu$m \\
\hline
Receiver radius & $R_v$ & 2.5 $\mu$m \\
\hline
Vessel length & $L$ & 4.0 mm \\
\hline
\multirow{2}{10em}{\centering Diffusion coefficient of information molecules} & $D_m$ & $9\times{10^{-9}}$ m$^2$/s \\
&  &  \\
\hline
Diffusion coefficient of TX & $D_{tx}$ & $2 \times {10^{ - 9}}$ m$^2$/s \\
\hline
Diffusion coefficient of RX & $D_{rx}$ &$1 \times {10^{-10}}$ m$^2$/s \\
\hline
Sampling interval & $T$ & 0.1ms \\
\hline
\end{tabular}
\end{center}
\end{table}

\begin{figure}[!t]
  \centering
  \includegraphics[width=0.45\textwidth]{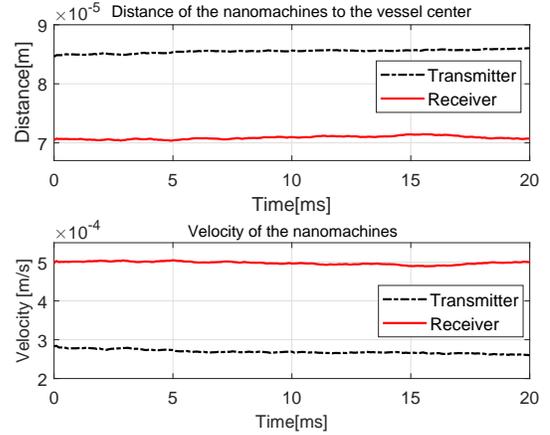}\\
  \caption{The distance to the center of the vessel and the corresponding velocity of the nanomachines. The distance and velocity of the terminals vary slowly but significantly.  Initial positions of the transmitter and receiver: $L_{tx}^0 = (0,6\cdot 10^{-5}\text{ m},6\cdot 10^{-5}\text{ m})$ and $L_{rx}^0 = (10\cdot 10^{-5}\text{ m},5\cdot 10^{-5}\text{ m},5\cdot 10^{-5}\text{ m})$, respectively.}  \label{Fig.4}
\end{figure}

%
%
In Fig. \ref{Fig.4}, we illustrate how the drift velocities of the transmitter and receiver change in an example trajectory.
From the figure,
we see that the location and drift velocity both vary with time, as indicated in (\ref{eqn:vnk}). While these changes are slow, they are significant and must be handled by the estimator.
Moreover, the figure illustrates that each of these nanomachines can experience different drift velocities as a result of their different positions in the nonuniform flow.

\begin{figure}[!t]
  \centering
  \includegraphics[width=0.45\textwidth]{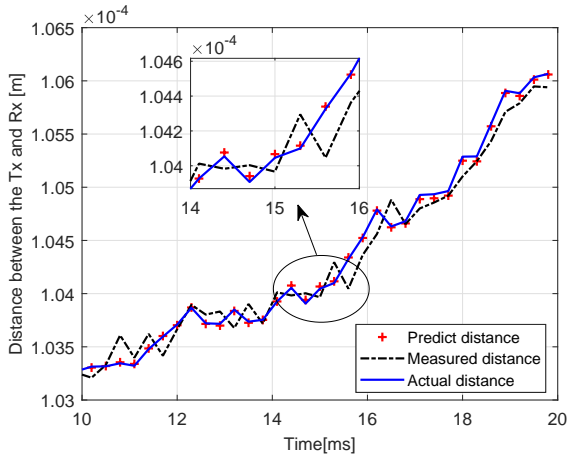}\\
   \caption{The comparison of true distance, measured distance and the extended Kalman filter predicted distance with feedback. The inset shows the high accuracy of the prediction compared with the measured distance. Note the changing slope of the distance in the inset figure, indicating a change in drift velocity. Initial positions of the transmitter and receiver are $L_{tx}^0 = (0,6\cdot 10^{-5}\text{ m},6\cdot 10^{-5}\text{ m})$ and $L_{rx}^0 = (10\cdot 10^{-5}\text{ m},5\cdot 10^{-5}\text{ m},5\cdot 10^{-5}\text{ m})$, respectively.}  \label{Fig6}
\end{figure}

In Fig. \ref{Fig6}, we illustrate the performance of the extended Kalman filter  with feedback.
For comparison, we use the distance derived from the noisy position estimates of each terminal, without using the Kalman filter. These position estimates are given by (\ref{eqn:L-predict}), and we call the resulting distance the {\em measured distance}. Note that these same position estimates are fed to the extended Kalman filter, and the distance estimates from the filter are called the {\em predicted distance}. We also compare with the ground truth {\em actual distance}.
As can be seen from Fig. \ref{Fig6}, the predicted distance is much closer to the true distance than the measured distance. It is interesting to note in the figure that the predictor correctly tracks the actual distance, even though the slope of the distance curve changes, indicating a change in drift velocity.

\begin{figure}[!t]
  \centering
  \includegraphics[width=0.45\textwidth]{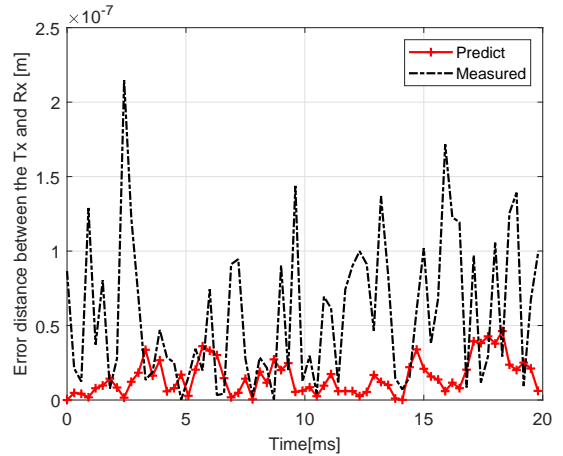}\\
  \caption{Comparison of measured error and extend Kalman filter prediction error. Initial positions are the same as in Figure \ref{Fig6}.}   \label{Fig7}
\end{figure}

To further illustrate the accuracy of the algorithm,
the measurement error (i.e., difference to the actual value) for the measured and predicted distances are compared in Fig. \ref{Fig7}. We can see clearly that the extended Kalman filter prediction error is much lower than the attenuation-based measured distance. In the simulation results, the mean of the measured distance error and the extended Kalman filter prediction error are $0.24473$ $\mu$m and $0.082697$ $\mu$m,
respectively. The amplitude of fluctuation of the measured error is roughly $\left(0,1.5\right)$ $\mu$m, while that of the prediction error is less than $\left(0,0.3\right)$ $\mu$m. These results confirm our expectations of strong performance from the extended Kalman filter.

\begin{figure}[!t]
  \centering
  \includegraphics[width=0.45\textwidth]{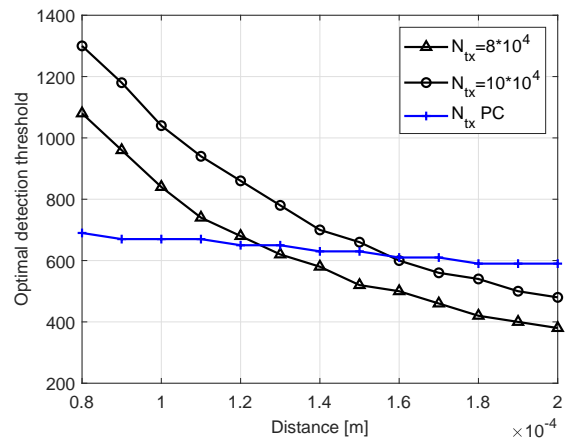}\\
  \caption{The optimal detection threshold versus distance between the transmitter and the receiver, for power control (PC) and non-PC.} \label{Threshold_distance}
\end{figure}

In the remaining figures, we analyze the optimal detection threshold and the performance of power control.
In Fig. \ref{Threshold_distance}, we show how the optimal detection threshold varies with distance under different number of transmitted molecules and power control scheme. For a given distance, the larger number of transmitted molecules, the optimal detection threshold is also larger. Moreover, as the distance increases, the optimal detection threshold decreases, but the decrease while using power control is much smaller. Therefore, in the mobile MC, for a constant number of transmitted molecules, calculation of the optimal detection threshold should be performed in every step to mitigate ISI, while for the power control scheme, a pre-set optimal detection threshold can achieve good performance.

\begin{figure}[!t]
  \centering
  \includegraphics[width=0.45\textwidth]{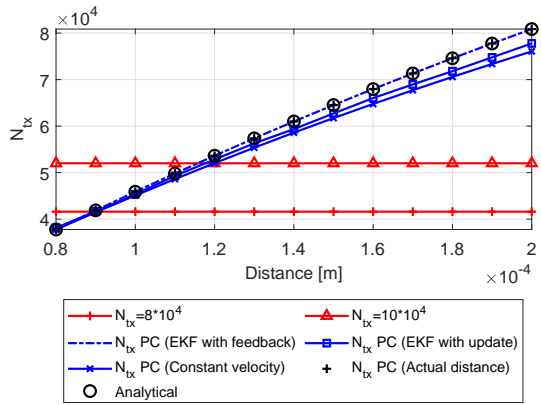}\\
  \caption{ The number of transmitted molecules per bit versus distance between the transmitter and the receiver, for power control (PC) and non-PC. The initial positions of the transmitter and receiver are $L_{tx}^0 = (0,6\cdot 10^{-5}\text{ m},6\cdot 10^{-5}\text{ m})$ and $L_{rx}^0 = (7.5\cdot 10^{-5}\text{ m},4\cdot 10^{-5}\text{ m},4\cdot 10^{-5}\text{ m})$, respectively.
  }\label{Fig11}
\end{figure}

\begin{figure}[!t]
  \centering
  \includegraphics[width=0.45\textwidth]{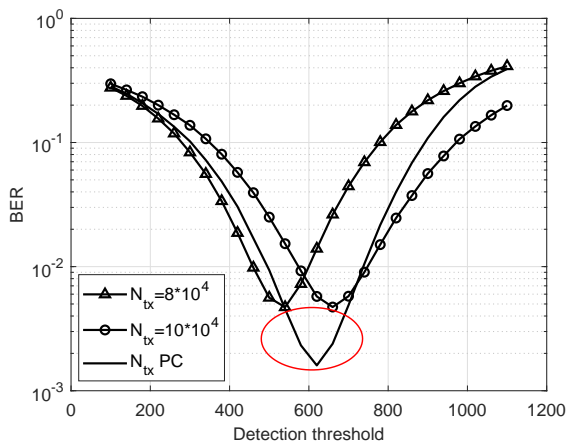}\\
  \caption{The BER versus detection threshold under different number of transmitted molecules and power control (PC) given a distance of $d_x = 1.5 \cdot 10^{-4}$m.} \label{BER_threshold}
\end{figure}

In Figures \ref{Fig11} and \ref{BER_threshold}, we illustrate the performance of power control using the optimal threshold.
In Fig. \ref{Fig11}, we illustrate the effect of power control by comparing the mean number of transmitted molecules with distance for one bit. For comparison, we illustrate constant values of $N_{tx}$ of $N_{tx}=8 \cdot 10^4$, $N_{tx}=10 \cdot 10^4$.
Meanwhile, in Figure \ref{BER_threshold}, we consider how the BER varies with detection threshold, both with and without power control, at a distance of $d_x=1.5 \cdot 10^{-4}$ m, and using the same constant $N_{tx}$ values as in Figure \ref{Fig11}. We can see from the simulation result, the BER of our power control scheme is better than that of the constant number of transmitted molecules schemes at their corresponding optimal detection threshold, specifically $5 \cdot 10^{-3}$, emphasizing the utility of optimal power control at each distance.
Though the optimal detection threshold for power control scheme shown in Fig. \ref{Threshold_distance} varies slightly with distance, as shown in the red circle of Fig. \ref{BER_threshold}, the performance of the power control scheme still outperforms the best performance when a constant number of transmitted molecules is used. Therefore, the simulation result proved the feasibility of a optimal detection threshold with the variation of distance for power control scheme.

\begin{figure}[!t]
  \centering
  \includegraphics[width=0.45\textwidth]{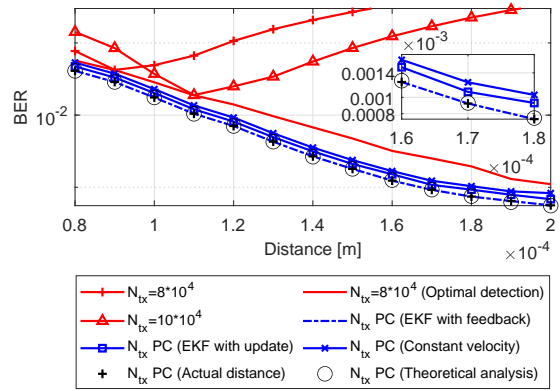}\\
  \caption{The BER versus distance between the transmitter and the receiver, for power control (PC) and non-PC. The initial positions of the nanomachines are the same as in Figure \ref{Fig11}.}\label{Fig12}
\end{figure}

In Fig. \ref{Fig12}, we show how the BER varies with distance under different numbers of transmitted molecules for signal-to-noise ratio (SNR) of 15, which is defined as the ratio of the number of received molecules $N_{rx}$ and the Gaussian counting noise variance $\sigma_n^2$ at the receiver. We compare BER using both $N_{tx}=8 \cdot 10^4$ and $N_{tx}=10 \cdot 10^4$ using an optimal detection threshold (i.e. the optimal detection threshold is calculated in every step as a function of distance), and $N_{tx}$ PC schemes. For $N_{tx}$ PC, we compared the power control schemes based on the extended Kalman filter distance and the actual distance. We can see that the power control scheme achieves near-optimal performance with feedback, and good performance with update.

Considering BCSK with a constant number of transmitted molecules, the best performance is achieved at a particular distance: this is because, for a given transmission power and detection threshold, the optimal distance is unique. Meanwhile, the power control scheme takes advantage of the residual molecules from the previous bit interval and reduces the molecules that accumulate in the channel, so that the performance is better than BCSK with an optimal detection threshold. However, for BCSK with power control, the ISI increases as the distance decreases,
so the performance is better as distance increases.
This is reflected in the performance observed in Figure \ref{Fig12}.

\section{conclusion}
In this paper, we considered a mobile MC system, and analyzed that system with a realistic blood-vessel-type propagation model, in which the transmitter and receiver experience different flow velocity.
An extended Kalman filter was employed to predict the changing distance between the transmitter and the receiver.
Based on the predicted distance, to keep the number of received molecules for bit 1 at a stable level, we performed power control to reduce the residual molecules in the channel to mitigate ISI. We also verified the effectiveness of a constant detection threshold for mobile MC by the proposed power control scheme. Finally, we analyzed the BER performance of the mobile MC system.
Simulation results verified the effectiveness of distance prediction and power control in mobile MC systems.

\appendix

\section{}
The mean of the received molecules under hypothesis $H_0$ during the $k$th time slot $\mu_{0,k}$ can be expressed as:
\begin{align}
\label{equation 73}
{\mu _{0,k}} &= {\mu _{I,k}} + {\mu _{n,k}}.\\\nonumber
\end{align}
As $\mu _{I,k}$ is the mean number of received molecules from time slot $1$ to $k-1$, it can be expressed as $\mu _{I,k} = E\left\{ {\sum\limits_{i = 1}^{k - 1} {N_{rx,i}^I} } \right\}$. Meanwhile, $\mu_{n,k}$ is the mean of the counting noise at the receiver which follows a Gaussian distribution with zero mean and the variance $\sigma_{0,k}^2$ depends on the average number of received molecules. Therefore, (\ref{equation 73}) can be expressed as

\begin{align}
\begin{split}
\label{equation 74}
{\mu _{0,k}} &= E\left\{ {\sum\limits_{i = 1}^{k - 1} {N_{rx,i}^I} } \right\}\\
 &= \sum\limits_{i = 1}^{k - 1} {E\left\{ {N_{rx,i}^I} \right\}},
\end{split}
\end{align}
and $E\left\{ {N_{rx,i}^I} \right\}$ can be expressed as
\begin{align}
\begin{split}
\label{equation 75}
E\left\{ {N_{rx,i}^I} \right\} &= {\Pr}\left( {b_{tx,i} = 1} \right)E\left\{ {N_{rx,i}^I|b_{tx,i} = 1} \right\} \\
&+ {\Pr}\left( {b_{tx,i} = 0} \right)E\left\{ {N_{rx,i}^I|b_{tx,i} = 0} \right\}\\
 &= {\Pr}\left( {b_{k - i} = 1} \right)\sum\limits_{i = 1}^{k - 1} {{N_{tx,i}}{P_{k-i+1}}}.
\end{split}
\end{align}
As we assume the same probability to transmit bit 0 and bit 1, which means $
{{\rm{P}}{\rm{r}}}\left( {b_{tx,k} = 0} \right) = {{\rm{P}}{\rm{r}}}\left( {b_{tx,k} = 1} \right) = \frac{1}{2}$, then
\begin{align}
\label{equation 76}
{\mu _{0,k}} = \frac{1}{2}\sum\limits_{i = 1}^{k - 1} {{N_{tx,i}}{P_{k-i+1}}}.
\end{align}
The variance of the received molecules under hypothesis $H_0$ during the $k$th time slot $\sigma_{0,k}^2$ can be expressed as:
\begin{align}
\begin{split}
\label{equation 77}
\sigma _{0,k}^2 = \sum\limits_{i = 1}^{k - 1} {\sigma _{I,i}^2}  + \sigma _{n,k}^2\\
 = \sum\limits_{i = 1}^{k - 1} {\sigma _{I,i}^2}  + {\mu _{0,k}},
\end{split}
\end{align}
and $\sigma _{I,j}^2$ can be expressed as
\begin{align}
\label{equation 78}
\sigma _{I,i}^2 = E\left\{ {{{\left( {N_{rx,i}^I} \right)}^2}} \right\} - {E^2}\left\{ {N_{rx,i}^I} \right\} .
\end{align}
As
\begin{align}
\label{equation 79}
{E^2}\left\{ {N_{rx,i}^I} \right\} = {\left( {\frac{1}{2}{N_{tx,i}}{P_{k-i+1}}} \right)^2} ,
\end{align}
and
\begin{align}
\label{equation 80}
\begin{split}
& E\left\{ {{{\left( {N_{rx,i}^I} \right)}^2}} \right\} = {\Pr}\left( {b_ {tx,i} = 1} \right)E\left( {{{\left( {N_{rx,i}^I} \right)}^2}|b_ {tx,i}  = 1} \right)\\
 &+ {\Pr}\left( {b_ {tx,i} = 0} \right)E\left( {{{\left( {N_{rx,i}^I} \right)}^2}|b_ {tx,i} = 0} \right)\\
 &= \frac{1}{2}\left[ {{N_{tx,i}}{P_{k-i+1}}\left( {1 - {P_{k-i+1}}} \right) + {{\left( {{N_{tx,i}}{P_{k-i+1}}} \right)}^2}} \right] ,
\end{split}
\end{align}
therefore,
\begin{align}
\label{equation 81}
\begin{split}
\sigma _{I,i}^2 &= \frac{1}{2}\left[ {{N_{tx,i}}{P_{k-i+1}}\left( {1 - {P_{k-i+1}}} \right) + {{\left( {{N_{tx,i}}{P_{k-i+1}}} \right)}^2}} \right] \\
&- {\left( {\frac{1}{2}{N_{tx,i}}{P_{k-i+1}}} \right)^2}\\
 &= \frac{1}{2}\left[ {{N_{tx,i}}{P_{k-i+1}}\left( {1 - {P_{k-i+1}}} \right)} \right] + \frac{1}{4}{\left( {{N_{tx,i}}{P_{k-i+1}}} \right)^2},
 \end{split}
\end{align}

\begin{align}
\begin{split}
\label{equation 82}
\sigma _{0,k}^2 &= \sum\limits_{i = 1}^{k - 1} \left[ \frac{1}{2}\left[ {{N_{tx,i}}{P_{k-i+1}}\left( {1 - {P_{k-i+1}}} \right)} \right] \right. \\
&\left. + \frac{1}{4}{{\left( {{N_{tx,i}}{P_{k-i+1}}} \right)}^2} \right]+ {\mu _{0,k}}   .
\end{split}
\end{align}
The mean of the received molecules under hypothesis $H_1$ during the $k$th time slot $\mu_{1,k}$ can be expressed as:
\begin{align}
\begin{split}
\label{equation 83}
{\mu _{1,k}} &= E\left( {N_{rx,k}^c + N_{rx,k}^I + {N_{n,k}}} \right)\\
 &= E\left( {N_{rx,k}^c} \right) + E\left( {N_{rx,k}^p} \right)\\
 &= {N_{tx,k}}{P_1} + \frac{1}{2}\sum\limits_{i = 1}^{k - 1} {{N_{tx, i}}{P_{k-i+1}}} .
 \end{split}
\end{align}
The variance of the received molecules under hypothesis $H_1$ during the $k$th time slot $\sigma_{1,k}^2$ can be expressed as:
\begin{align}
\label{equation 84}
\begin{split}
\sigma _{1,k}^2 &= {\sigma_{c,k} ^2} + \sum\limits_{i = 1}^{k - 1} {\sigma _{I,i}^2}  + \sigma _{n,k}^2\\
 &= {N_{tx,k}}{P_1}\left( {1 - {P_1}} \right) + \sum\limits_{i = 1}^{k - 1} \left[ \frac{1}{2}\left[ {{N_{tx,i}}{P_{k-i+1}}\left( {1 - {P_{k-i+1}}} \right)} \right] \right. \\  
 & \left. + \frac{1}{4}{{\left( {{N_{tx, i}}{P_{k-i+1}}} \right)}^2} \right]  + {\mu _{1,k}}.
 \end{split}
\end{align}

\section{}
Considering the same probability to transmit bit $0$ and bit $1$, and the fact
$Q\left( x \right) = \frac{1}{2}\rm{erfc}\left( {\frac{x}{{\sqrt 2 }}} \right)$, the error probability $P_e$ can be expressed as:
\begin{align}
\begin{split}
{P_e} &= \frac{1}{2}Q\left( {\sqrt {\frac{{{{\left( {{N_{th}} - {\mu _0}} \right)}^2}}}{{\delta _0^2}}} } \right) + \frac{1}{2}\left( {1 - Q\left( {\sqrt {\frac{{{{\left( {{N_{th}} - {\mu _1}} \right)}^2}}}{{\delta _1^2}}} } \right)} \right)\\
 &= \frac{1}{2} - \frac{1}{4}\left[ {{\rm{erf}}\left( {\frac{{\left( {{N_{th}}{\rm{ - }}{\mu _{\rm{0}}}} \right)}}{{\sqrt {\rm{2}} {\delta _{\rm{0}}}}}} \right){\rm{ - erf}}\left( {\frac{{\left( {{N_{th}}{\rm{ - }}{\mu _{\rm{1}}}} \right)}}{{\sqrt {\rm{2}} {\delta _{\rm{1}}}}}} \right)} \right].
\end{split}
\end{align}
The optimal detection threshold can be expressed as:
\begin{align}
{N_{th,opt}} = \mathop {\min }\limits_{{N_{th}}} \left\{ {{P_e}} \right\}.
\end{align}
Therefore, the optimal detection threshold $N_{th,opt}$ can be achieved by
${\raise0.7ex\hbox{${\partial {P_e}}$} \!\mathord{\left/
 {\vphantom {{\partial {P_e}} {\partial {N_{th,opt}}}}}\right.\kern-\nulldelimiterspace}
\!\lower0.7ex\hbox{${\partial {N_{th}}}$}} = 0$.

\begin{align}
\begin{split}
\label{equ68}
\frac{{\partial {P_e}}}{{\partial {N_{th}}}} &=  - \frac{1}{4}\left[ {\sqrt {\frac{2}{{\pi \sigma _0^2}}} \exp \left( {\frac{{{{\left( {{N_{th}} - {\mu _0}} \right)}^2}}}{{2\sigma _0^2}}} \right)} \right]\\ 
&+ \frac{1}{4}\left[ {\sqrt {\frac{2}{{\pi \sigma _1^2}}} \exp \left( {\frac{{{{\left( {{N_{th}} - {\mu _1}} \right)}^2}}}{{2\sigma _1^2}}} \right)} \right].
\end{split}
\end{align}
Letting the quantity in $\left( \ref{equ68}\right)$ be equal to zero, the equation can be expressed as:
\begin{align}
\begin{split}
\label{equ75}
 &- \frac{1}{4}\left[ {\sqrt {\frac{2}{{\pi \sigma _0^2}}} \exp \left( {\frac{{{{\left( {{N_{th}} - {\mu _0}} \right)}^2}}}{{2\sigma _0^2}}} \right)} \right]\\ 
 &+ \frac{1}{4}\left[ {\sqrt {\frac{2}{{\pi \sigma _1^2}}} \exp \left( {\frac{{{{\left( {{N_{th}} - {\mu _1}} \right)}^2}}}{{2\sigma _1^2}}} \right)} \right] = 0.\\ 
\end{split}
\end{align}
Then, $\left( \ref{equ75} \right)$ can be expressed as:
\begin{align}
\begin{split}
\label{equ76}
\left[ {\frac{1}{{{\sigma _0}}}\exp \left( {\frac{{{{\left( {{N_{th}} - {\mu _0}} \right)}^2}}}{{2\sigma _0^2}}} \right)} \right] = \left[ {\frac{1}{{{\sigma _1}}}\exp \left( {\frac{{{{\left( {{N_{th}} - {\mu _1}} \right)}^2}}}{{2\sigma _1^2}}} \right)} \right].
\end{split}
\end{align}
%
Taking the logarithm of both sides of $\left(\ref{equ76}\right)$,
\begin{align}
\frac{{{{\left( {{N_{th}} - {\mu _0}} \right)}^2}}}{{2\sigma _0^2}} - \frac{{{{\left( {{N_{th}} - {\mu _0}} \right)}^2}}}{{2\sigma _0^2}} - \ln \frac{{{\sigma _0}}}{{{\sigma _1}}} = 0 .
\end{align}
Therefore, the optimal detection threshold $N_{th}$ can be expressed as:
\begin{align}
\label{equ45}
{N_{th,opt}} = \frac{{{\mu _0}\sigma _1^2 - {\mu _1}\sigma _0^2{\rm{ + }}\sqrt {\left( {{\mu _0}\sigma _1^2{\rm{ - }}{\mu _1}\sigma _0^2} \right){\rm{ - }}\left( {\sigma _1^2 - \sigma _0^2} \right){\mathcal{C}}} }}{{\sigma _1^2 - \sigma _0^2}},
\end{align}
where
${\mathcal{C}} = \left( {{\mu _0}\sigma _1^2{\rm{ - }}{\mu _1}\sigma _0^2} \right) - 2\sigma _0^2\sigma _1^2 \cdot \ln \frac{{{\sigma _0}}}{{{\sigma _1}}}$.

\bibliographystyle{IEEEtran}
\bibliography{references}

\begin{thebibliography}{10}
\providecommand{\url}[1]{#1}
\csname url@samestyle\endcsname
\providecommand{\newblock}{\relax}
\providecommand{\bibinfo}[2]{#2}
\providecommand{\BIBentrySTDinterwordspacing}{\spaceskip=0pt\relax}
\providecommand{\BIBentryALTinterwordstretchfactor}{4}
\providecommand{\BIBentryALTinterwordspacing}{\spaceskip=\fontdimen2\font plus
\BIBentryALTinterwordstretchfactor\fontdimen3\font minus
  \fontdimen4\font\relax}
\providecommand{\BIBforeignlanguage}[2]{{%
\expandafter\ifx\csname l@#1\endcsname\relax
\typeout{** WARNING: IEEEtran.bst: No hyphenation pattern has been}%
\typeout{** loaded for the language `#1'. Using the pattern for}%
\typeout{** the default language instead.}%
\else
\language=\csname l@#1\endcsname
\fi
#2}}
\providecommand{\BIBdecl}{\relax}
\BIBdecl

\bibitem{farsad2016comprehensive}
N.~Farsad, H.~B. Yilmaz, A.~Eckford, C.-B. Chae, and W.~Guo, ``A comprehensive
  survey of recent advancements in molecular communication,'' \emph{IEEE
  Communications Surveys \& Tutorials}, vol.~18, no.~3, pp. 1887--1919, 2016.

\bibitem{akyildiz2019moving}
I.~F. Akyildiz, M.~Pierobon, and S.~Balasubramaniam, ``Moving forward with
  molecular communication: from theory to human health applications,''
  \emph{Proceedings of the IEEE}, vol. 107, no.~5, pp. 858--865, 2019.

\bibitem{chen2021resource}
X.~Chen, M.~Wen, C.-B. Chae, L.-L. Yang, F.~Ji, and K.~K. Igorevich, ``Resource
  allocation for multi-user molecular communication systems oriented to the
  internet of medical things,'' \emph{IEEE Internet of Things Journal}, 2021.

\bibitem{khalid2020modeling}
M.~Khalid, O.~Amin, S.~Ahmed, B.~Shihada, and M.-S. Alouini, ``Modeling of
  viral aerosol transmission and detection,'' \emph{IEEE Transactions on
  Communications}, vol.~68, no.~8, pp. 4859--4873, 2020.

\bibitem{nakano2019methods}
T.~Nakano, Y.~Okaie, S.~Kobayashi, T.~Hara, Y.~Hiraoka, and T.~Haraguchi,
  ``Methods and applications of mobile molecular communication,''
  \emph{Proceedings of the IEEE}, vol. 107, no.~7, pp. 1442--1456, 2019.

\bibitem{varshney2018flow}
N.~Varshney, W.~Haselmayr, and W.~Guo, ``On flow-induced diffusive mobile
  molecular communication: First hitting time and performance analysis,''
  \emph{IEEE Transactions on Molecular, Biological and Multi-Scale
  Communications}, vol.~4, no.~4, pp. 195--207, 2018.

\bibitem{srinivas2012molecular}
K.~V. Srinivas, A.~W. Eckford, and R.~S. Adve, ``Molecular communication in
  fluid media: The additive inverse gaussian noise channel,'' \emph{IEEE
  transactions on information theory}, vol.~58, no.~7, pp. 4678--4692, 2012.

\bibitem{lin2019concentration}
L.~Lin, Q.~Wu, M.~Ma, and H.~Yan, ``Concentration-based demodulation scheme for
  mobile receiver in molecular communication,'' \emph{Nano Communication
  Networks}, vol.~20, pp. 11--19, 2019.

\bibitem{chouhan2019optimal}
L.~Chouhan, P.~K. Sharma, and N.~Varshney, ``Optimal transmitted molecules and
  decision threshold for drift-induced diffusive molecular channel with mobile
  nanomachines,'' \emph{IEEE transactions on nanobioscience}, vol.~18, no.~4,
  pp. 651--660, 2019.

\bibitem{huang2019statistical}
S.~Huang, L.~Lin, H.~Yan, J.~Xu, and F.~Liu, ``Statistical analysis of received
  signal and error performance for mobile molecular communication,'' \emph{IEEE
  transactions on nanobioscience}, vol.~18, no.~3, pp. 415--427, 2019.

\bibitem{cao2019diffusive}
T.~N. Cao, A.~Ahmadzadeh, V.~Jamali, W.~Wicke, P.~L. Yeoh, J.~Evans, and
  R.~Schober, ``Diffusive mobile mc with absorbing receivers: Stochastic
  analysis and applications,'' \emph{IEEE Transactions on Molecular, Biological
  and Multi-Scale Communications}, vol.~5, no.~2, pp. 84--99, 2019.

\bibitem{gursoy2021concentration}
M.~C. Gursoy, D.~Seo, and U.~Mitra, ``A concentration-time hybrid modulation
  scheme for molecular communications,'' \emph{IEEE Transactions on Molecular,
  Biological and Multi-Scale Communications}, 2021.

\bibitem{kislal2019isi}
A.~O. Kislal, H.~B. Yilmaz, A.~E. Pusane, and T.~Tugcu, ``Isi-aware channel
  code design for molecular communication via diffusion,'' \emph{IEEE
  transactions on nanobioscience}, vol.~18, no.~2, pp. 205--213, 2019.

\bibitem{li2019csi}
B.~Li, W.~Guo, X.~Wang, Y.~Deng, Y.~Lan, C.~Zhao, and A.~Nallanathan,
  ``Csi-independent non-linear signal detection in molecular communications,''
  \emph{IEEE Transactions on Signal Processing}, vol.~68, pp. 97--112, 2019.

\bibitem{thakur2020iterative}
M.~S. Thakur, S.~Sharma, and V.~Bhatia, ``Iterative signal detection for
  diffusion-based molecular communications,'' \emph{IEEE Transactions on
  Molecular, Biological and Multi-Scale Communications}, vol.~6, no.~1, pp.
  36--49, 2020.

\bibitem{qian2021k}
X.~Qian, M.~Di~Renzo, and A.~Eckford, ``K-means clustering-aided non-coherent
  detection for molecular communications,'' \emph{IEEE Transactions on
  Communications}, 2021.

\bibitem{chen2020generalized}
X.~Chen, Y.~Huang, L.-L. Yang, and M.~Wen, ``Generalized molecular-shift keying
  (gmosk): Principles and performance analysis,'' \emph{IEEE Transactions on
  Molecular, Biological and Multi-Scale Communications}, vol.~6, no.~3, pp.
  168--183, 2020.

\bibitem{tang2020molecular}
Y.~Tang, M.~Wen, X.~Chen, Y.~Huang, and L.-L. Yang, ``Molecular type
  permutation shift keying for molecular communication,'' \emph{IEEE
  Transactions on Molecular, Biological and Multi-Scale Communications},
  vol.~6, no.~2, pp. 160--164, 2020.

\bibitem{ahmadzadeh2018stochastic}
A.~Ahmadzadeh, V.~Jamali, and R.~Schober, ``Stochastic channel modeling for
  diffusive mobile molecular communication systems,'' \emph{IEEE Transactions
  on Communications}, vol.~66, no.~12, pp. 6205--6220, 2018.

\bibitem{tepekule2015isi}
B.~Tepekule, A.~E. Pusane, H.~B. Yilmaz, C.-B. Chae, and T.~Tugcu, ``Isi
  mitigation techniques in molecular communication,'' \emph{IEEE Transactions
  on Molecular, Biological and Multi-Scale Communications}, vol.~1, no.~2, pp.
  202--216, 2015.

\bibitem{jing2020power}
D.~Jing, Y.~Li, and A.~W. Eckford, ``Power control for isi mitigation in mobile
  molecular communication,'' \emph{IEEE Communications Letters}, 2020.

\bibitem{huang2013distance}
J.-T. Huang, H.-Y. Lai, Y.-C. Lee, C.-H. Lee, and P.-C. Yeh, ``Distance
  estimation in concentration-based molecular communications,'' in \emph{2013
  IEEE Global Communications Conference (GLOBECOM)}.\hskip 1em plus 0.5em minus
  0.4em\relax IEEE, 2013, pp. 2587--2591.

\bibitem{wang2015distance}
X.~Wang, M.~D. Higgins, and M.~S. Leeson, ``Distance estimation schemes for
  diffusion based molecular communication systems,'' \emph{IEEE Communications
  Letters}, vol.~19, no.~3, pp. 399--402, 2015.

\bibitem{lin2016parameter}
L.~Lin, C.~Yang, S.~Ma, and M.~Ma, ``Parameter estimation of inverse gaussian
  channel for diffusion-based molecular communication,'' in \emph{2016 IEEE
  wireless communications and networking conference}.\hskip 1em plus 0.5em
  minus 0.4em\relax IEEE, 2016, pp. 1--6.

\bibitem{chen2020parameter}
Y.~Chen, Y.~Li, L.~Lin, and H.~Yan, ``Parameter estimation of diffusive
  molecular communication with drift,'' \emph{IEEE Access}, vol.~8, pp.
  142\,704--142\,713, 2020.

\bibitem{huang2020initial}
S.~Huang, L.~Lin, W.~Guo, H.~Yan, J.~Xu, and F.~Liu, ``Initial distance
  estimation and signal detection for diffusive mobile molecular
  communication,'' \emph{IEEE Transactions on NanoBioscience}, 2020.

\bibitem{moore2012measuring}
M.~J. Moore, T.~Nakano, A.~Enomoto, and T.~Suda, ``Measuring distance from
  single spike feedback signals in molecular communication,'' \emph{IEEE
  Transactions on Signal Processing}, vol.~60, no.~7, pp. 3576--3587, 2012.

\bibitem{turan2018transmitter}
M.~Turan, B.~C. Akdeniz, M.~{\c{S}}. Kuran, H.~B. Yilmaz, I.~Demirkol, A.~E.
  Pusane, and T.~Tugcu, ``Transmitter localization in vessel-like diffusive
  channels using ring-shaped molecular receivers,'' \emph{IEEE Communications
  Letters}, vol.~22, no.~12, pp. 2511--2514, 2018.

\bibitem{liu2019localization}
S.~Liu, S.~Bao, and C.~Zhao, ``Localization schemes for 2-d molecular
  communication via diffusion,'' in \emph{International Conference in
  Communications, Signal Processing, and Systems}.\hskip 1em plus 0.5em minus
  0.4em\relax Springer, 2019, pp. 749--756.

\bibitem{kumar2020nanomachine}
S.~Kumar, ``Nanomachine localization in a diffusive molecular communication
  system,'' \emph{IEEE Systems Journal}, vol.~14, no.~2, pp. 3011--3014, 2020.

\bibitem{yetimoglu2021multiple}
O.~Yetimoglu, M.~K. Avci, B.~C. Akdeniz, H.~B. Yilmaz, A.~E. Pusane, and
  T.~Tugcu, ``Multiple transmitter localization via single receiver in 3-d
  molecular communication via diffusion,'' \emph{Digital Signal Processing}, p.
  103185, 2021.

\bibitem{wicke2018modeling}
W.~Wicke, T.~Schwering, A.~Ahmadzadeh, V.~Jamali, A.~Noel, and R.~Schober,
  ``Modeling duct flow for molecular communication,'' in \emph{2018 IEEE Global
  Communications Conference (GLOBECOM)}.\hskip 1em plus 0.5em minus 0.4em\relax
  IEEE, 2018, pp. 206--212.

\bibitem{felicetti2014molecular}
L.~Felicetti, M.~Femminella, G.~Reali, and P.~Li{\`o}, ``A molecular
  communication system in blood vessels for tumor detection,'' in
  \emph{Proceedings of ACM The First Annual International Conference on
  Nanoscale Computing and Communication}, 2014, pp. 1--9.

\bibitem{he2016channel}
P.~He, Y.~Mao, Q.~Liu, P.~Li{\`o}, and K.~Yang, ``Channel modelling of
  molecular communications across blood vessels and nerves,'' in \emph{2016
  IEEE International Conference on Communications (ICC)}.\hskip 1em plus 0.5em
  minus 0.4em\relax IEEE, 2016, pp. 1--6.

\bibitem{batchelor2000introduction}
C.~K. Batchelor and G.~Batchelor, \emph{An introduction to fluid
  dynamics}.\hskip 1em plus 0.5em minus 0.4em\relax Cambridge university press,
  2000.

\bibitem{jamali2019channel}
V.~Jamali, A.~Ahmadzadeh, W.~Wicke, A.~Noel, and R.~Schober, ``Channel modeling
  for diffusive molecular communication—a tutorial review,''
  \emph{Proceedings of the IEEE}, vol. 107, no.~7, pp. 1256--1301, 2019.

\bibitem{bruus2008theoretical}
H.~Bruus, \emph{Theoretical microfluidics}.\hskip 1em plus 0.5em minus
  0.4em\relax Oxford university press Oxford, 2008, vol.~18.

\bibitem{wang2017highly}
R.~Wang, W.~Guo, X.~Li, Z.~Liu, H.~Liu, and S.~Ding, ``Highly efficient
  mof-based self-propelled micromotors for water purification,'' \emph{RSC
  advances}, vol.~7, no.~67, pp. 42\,462--42\,467, 2017.

\bibitem{wu2019microrobotic}
Z.~Wu, L.~Li, Y.~Yang, P.~Hu, Y.~Li, S.-Y. Yang, L.~V. Wang, and W.~Gao, ``A
  microrobotic system guided by photoacoustic computed tomography for targeted
  navigation in intestines in vivo,'' \emph{Science robotics}, vol.~4, no.~32,
  2019.

\bibitem{zhang2019micro}
Y.~Zhang, K.~Yuan, and L.~Zhang, ``Micro/nanomachines: from functionalization
  to sensing and removal,'' \emph{Advanced Materials Technologies}, vol.~4,
  no.~4, p. 1800636, 2019.

\bibitem{wang2018recent}
B.~Wang, Y.~Zhang, and L.~Zhang, ``Recent progress on micro-and nano-robots:
  Towards in vivo tracking and localization,'' \emph{Quantitative imaging in
  medicine and surgery}, vol.~8, no.~5, p. 461, 2018.

\bibitem{kim2018artificial}
K.~Kim, J.~Guo, Z.~Liang, and D.~Fan, ``Artificial micro/nanomachines for
  bioapplications: biochemical delivery and diagnostic sensing,''
  \emph{Advanced Functional Materials}, vol.~28, no.~25, p. 1705867, 2018.

\bibitem{yan2017multifunctional}
X.~Yan, Q.~Zhou, M.~Vincent, Y.~Deng, J.~Yu, J.~Xu, T.~Xu, T.~Tang, L.~Bian,
  Y.-X.~J. Wang \emph{et~al.}, ``Multifunctional biohybrid magnetite
  microrobots for imaging-guided therapy,'' \emph{Science Robotics}, vol.~2,
  no.~12, 2017.

\bibitem{moore2008molecular}
M.~J. Moore, A.~Enomoto, T.~Suda, A.~Kayasuga, and K.~Oiwa, ``Molecular
  communication: uni-cast communication on a microtubule topology,'' in
  \emph{2008 IEEE International Conference on Systems, Man and
  Cybernetics}.\hskip 1em plus 0.5em minus 0.4em\relax IEEE, 2008, pp. 18--23.

\bibitem{enomoto2011design}
A.~Enomoto, M.~J. Moore, T.~Suda, and K.~Oiwa, ``Design of self-organizing
  microtubule networks for molecular communication,'' \emph{Nano Communication
  Networks}, vol.~2, no.~1, pp. 16--24, 2011.

\bibitem{jiang2019microtubule}
R.~Jiang, S.~Vandal, S.~Park, S.~Majd, E.~T{\"u}zel, and W.~O. Hancock,
  ``Microtubule binding kinetics of membrane-bound kinesin-1 predicts high
  motor copy numbers on intracellular cargo,'' \emph{Proceedings of the
  National Academy of Sciences}, vol. 116, no.~52, pp. 26\,564--26\,570, 2019.

\bibitem{jeune2010cargo}
Y.~Jeune-Smith, A.~Agarwal, and H.~Hess, ``Cargo loading onto kinesin powered
  molecular shuttles,'' \emph{JoVE (Journal of Visualized Experiments)},
  no.~45, p. e2006, 2010.

\bibitem{reif1999stochastic}
K.~Reif, S.~Gunther, E.~Yaz, and R.~Unbehauen, ``Stochastic stability of the
  discrete-time extended kalman filter,'' \emph{IEEE Transactions on Automatic
  control}, vol.~44, no.~4, pp. 714--728, 1999.

\bibitem{daum2005nonlinear}
F.~Daum, ``Nonlinear filters: beyond the kalman filter,'' \emph{IEEE Aerospace
  and Electronic Systems Magazine}, vol.~20, no.~8, pp. 57--69, 2005.

\bibitem{9691339}
E.~Aslan, M.~E. Çelebi, and F.~Pekergin, ``Wiener and kalman detection methods
  for molecular communications,'' \emph{IEEE Transactions on NanoBioscience},
  pp. 1--1, 2022.

\bibitem{shi2020nanorobots}
S.~Shi, Y.~Yan, J.~Xiong, U.~K. Cheang, X.~Yao, and Y.~Chen,
  ``Nanorobots-assisted natural computation for multifocal tumor sensitization
  and targeting,'' \emph{IEEE Transactions on NanoBioscience}, vol.~20, no.~2,
  pp. 154--165, 2020.

\bibitem{li2020progress}
M.~Li, N.~Xi, Y.~Wang, and L.~Liu, ``Progress in nanorobotics for advancing
  biomedicine,'' \emph{IEEE Transactions on Biomedical Engineering}, vol.~68,
  no.~1, pp. 130--147, 2020.

\bibitem{kuran2010energy}
M.~{\c{S}}. Kuran, H.~B. Yilmaz, T.~Tugcu, and B.~{\"O}zerman, ``Energy model
  for communication via diffusion in nanonetworks,'' \emph{Nano Communication
  Networks}, vol.~1, no.~2, pp. 86--95, 2010.

\bibitem{pierobon2011diffusion}
M.~Pierobon and I.~F. Akyildiz, ``Diffusion-based noise analysis for molecular
  communication in nanonetworks,'' \emph{IEEE Transactions on signal
  processing}, vol.~59, no.~6, pp. 2532--2547, 2011.

\bibitem{kilinc2013receiver}
D.~Kilinc and O.~B. Akan, ``Receiver design for molecular communication,''
  \emph{IEEE Journal on Selected Areas in Communications}, vol.~31, no.~12, pp.
  705--714, 2013.

\end{thebibliography}
\end{document}